\documentclass[12pt]{article}
\usepackage[super,compress]{cite}
\usepackage{amsmath}
\usepackage{amssymb}
\usepackage{graphicx}
\usepackage{float}
\begin{document}

\begin{center}
{\bf Reaction and interaction nucleus-nucleus cross sections in the complete
Glauber theory }

Yu.M. Shabelski and A.G. Shuvaev \\

\vspace{.5cm}

NRC "Kurchatov Institute" - PNPI, Gatchina 188300 Russia\\

\vskip 0.9 truecm
E-mail: shabelsk@thd.pnpi.spb.ru\\
E-mail: shuvaev@thd.pnpi.spb.ru

\end{center}

\vspace{1.2cm}

\begin{abstract}
\noindent
The straightforward calculations of the reaction and interaction
cross sections of the nuclear scattering are carried out
in Glauber approach using the generating function method.
It allows for the resummation of all orders of Glauber theory.
The results are obtained for $^4$He, $^{11}$Li, $^{12}$C
scattering on $^{12}$C target.
The difference between the reaction and the differential cross
section is shown to be not exceeding several percents
\end{abstract}

\section{Introduction}

The information on the various aspects of the nuclear structure,
in particular, about halo nuclei, comes mainly from
the experimental data on the collision of the nucleus under study
$A$ with a target $B$.
What is directly measured in this reaction is
the interaction cross section.
The interaction cross section is defined as that
of the process when the beam nucleus $A$ scatters
without being excited or disintegrated whereas it is
allowed for the target nucleus $B$,
$$
\sigma_{AB}^{I}=\sigma_{AB}^{tot} -\sigma_{AB}^{el}
- \sigma_{AB\to AB^*}\,=\,
\sigma_{AB}^{tot} -\sigma_{AB\to AB^\prime}.
$$
Here $B^*$ stands for all the excited or disintegrated states
of the target nucleus, $B^\prime=\{B, B^*\}$ denotes the complete
set of the target states.
Introducing the reaction, or the total inelastic, cross section
$$
\sigma_{AB}^{R} = \sigma_{AB}^{tot} -\sigma_{AB}^{el},
$$
it can be rewritten as
$$
\sigma_{AB}^{I} = \sigma_{AB}^{R} - \sigma_{AB\to AB^*}
$$
The reason why it is the interaction cross section
that is really measured is in the experimental difficulty
to distinguish the pure elastic scattering
from the processes giving rise to the target excitation
or disintegration.
The beam energy loss in the latter case is very
small compared to the initial value to detect.

Usually the difference between $\sigma_{AB}^{R}$ and $\sigma_{AB}^{I}$
is assumed to be negligible. The Monte-Carlo simulation results into
the value 2-3\% of $\sigma_{AB}^{R}$~\cite{Novikov:2013zdw}.
In this paper we present a complete analytical Glauber calculation
of the reaction and the interaction cross sections for several
relatively light nuclei, $^4$He, $^{11}$Li, $^{12}$C.


\section{Complete Glauber calculation of the interaction
cross section}

The amplitude of the elastic scattering of the incident
nucleus $A$ on the fixed target nucleus $B$ reads
in the Glauber theory
\begin{equation}
\label{bint}
f_{AB}(q)\,=\,\frac{ip}{2\pi}\int d^{2}b\,e^{i qb}\,
\bigl[1\,-\,s_{AB}(b)\bigr].
\end{equation}
Here $p$ is a relative momentum in the central of mass frame,
$q$ is the transferred momentum. The impact parameter $b$ is
a two dimensional vector
in the transverse plane with respect to relative momentum
of the colliding nuclei $A$, $B$.
The evaluation of the function $s_{AB}(b)$
relies on the short range of the strong
interaction.
Due to this property the phase shift on a nucleus
comes out the sum of those for the independent
scattering of the constituent nucleons.
The function $s_{AB}(b)$ reads
\begin{equation}
\label{sab}
s_{AB}(b)\,=\,\langle\,A,\,|\langle\,B\,|
\left\{\prod\limits_{i\,j}\bigl[1-\Gamma_{NN}(b+x_i-y_j)\bigr]
\right\}
|\,A,\,|\rangle\,B\,\rangle,
\end{equation}
where
$$
\Gamma_{NN}(b)\,\equiv\,
1\,-\,s_{NN}(b)\,=\,\frac 1{2\pi i p}\int d^2q\,e^{iqb}f_{NN}(q),
$$
$f_{NN}(q)$ and $s_{NN}(b)$ are the nucleon-nucleon elastic scattering
amplitude and the phase shift. The brackets
stand for an average over the nucleons' positions
$x_i$ and $y_j$ lying in the same plain with the impact
parameter. Each pair
$\{i,j\}$ enters the product only once, meaning that
each nucleon from the projectile nucleus can scatter
on each nucleon from the target no more than once.

The elastic nucleon-nucleon amplitude, $f_{NN}$, is mainly
imaginary at the beam energy about 1~GeV per nucleon,
$\mathrm{Re} f_{NN}/\mathrm{Im} f_{NN}\lesssim 10^{-1}$.
The standard parametrization is
\begin{equation}
\label{fNN}
f_{NN}(q)\,=\,ip\frac{\sigma_{NN}^{tot}}{4\pi}
e^{-\frac 12\beta q^2},
\end{equation}
where $\sigma_{NN}^{tot}$ is the total nucleon-nucleon
cross section. The slope $\beta$ is related to an effective
interaction radius $a^2 = 2\pi\beta$.

The elastic amplitude is simple related to the total cross section
through the optic theorem,
$$
\sigma_{AB}^{tot}\,=\,\frac{4\pi}{p}
\mathrm{Im}f_{AB}(q=0)\,=\,
2\int\!\! d^2b\,\bigl[1\,-\,s_{AB}(b)\bigr].
$$
The difference between the total cross section
and the integrated elastic cross section,
\begin{equation}
\label{elastic}
\sigma_{AB}^{el}\,=\,
\int\!\! d^2b\,\bigl[1\,-\,s_{AB}(b)\bigr]^2,
\end{equation}
yields the reaction cross section,
$$
\sigma_{AB}^{r}\,=\,\sigma_{AB}^{tot}\,-\,\sigma_{AB}^{el}
\,=\,
\int\!\! d^2b\,\bigl[1\,-\,s_{AB}^2(b)\bigr].
$$

The amplitude of the process when
the target nucleus $B$ is exited or disintegrated after collision
with the projectile takes in the Glauber approach a form similar
to (\ref{sab}),
\begin{equation}
\label{sab*}
s_{AB^{*}}(b)\,=\,\langle\,A,\,B^{*}\,|
\left\{\prod\limits_{i\,j}\bigl[1-\Gamma_{NN}(b+x_i-y_j)\bigr]
\right\}
|\,A,\,B\,\rangle,
\end{equation}
Denoting through
$|B^{\prime}\,\rangle = \{|B\,\rangle, |B^{*}\,\rangle \}$
the set of all target states and using its completeness,
\begin{equation}
\label{complete}
\sum_{B^{\prime}}|B^{\prime}\,\rangle\,\langle B^{\prime}\,|
\,=\,|B\,\rangle\,\langle B\,|
\,+\,\sum_{B^{*}}|B^{*}\,\rangle\,\langle B^{*}\,|\,=\,1,
\end{equation}
one gets for the cross section $A B \to A B^\prime$
\begin{equation}
\label{ABprim}
\sigma_{A B \to A B^\prime}\,=\,\int d^{2}b\,
\bigl[\,1\,-\,2s_{AB}(b)\,+\,J_{AB}(b)\bigr],
\end{equation}
where
\begin{eqnarray}
\label{JAB}
J_{AB}(b)\,&=&\,\langle\,A\,|\langle\,B\,|
\left\{\prod\limits_{i\,j}\bigl[1-\Gamma_{NN}(b+x_i-y_j)\bigr]
\right\}
|\,A\,\rangle  \\
&&\times
\langle\,A\,|
\left\{\prod\limits_{i\,j^\prime}
\bigl[1-\Gamma_{NN}(b+x_i-y_j^\prime)\bigr]
\right\}
|\,B\,\rangle\,|\,A\,\rangle. \nonumber
\end{eqnarray}
Subtracting from (\ref{ABprim}) the elastic cross
section~(\ref{elastic})
we arrive at the interaction cross section,
\begin{equation}
\label{interact}
\sigma_{AB}^{I}\,=\,\sigma_{AB}^{r}\,-\,\sigma_{A B \to A B^*},
~~~~
\sigma_{A B \to A B^*}\,=\,\int d^{2}b\,
\bigl[\,J_{AB}(b)\,-\,s_{AB}^2(b)\,\bigr].
\end{equation}

The generating function method \cite{Shabelski:2021iqk} relies on the identity
\begin{eqnarray}
&&
\int \frac{D\Phi D\Phi^*}{2\pi i}\exp\biggl\{
-\int d^{\,2}x d^{\,2}y\,\Phi(x)\Delta^{-1}(x-y)\Phi^*(y)
+\sum_i\Phi(x_i)+\sum_j\Phi^*(y_j)
\biggr\} \nonumber \\
\label{GI}
&=&\,\exp\biggl\{\sum\limits_{i,j}\Delta(x_i-y_j)
\biggr\}\,=\,
\left\{\prod\limits_{i\,j}\bigl[1-\Gamma_{NN}(x_i-y_j)\bigr]
\right\},
\end{eqnarray}
valid for the function $\Delta(x-y)$ chosen to obey
the equation
\begin{equation}
\label{Delta}
e^{\,\Delta(x-y)}\,-\,1\,=\,-\Gamma_{NN}(x-y).
\end{equation}
The functional integral can be thought of as
an infinite product of two dimensional integrals
over the auxiliary independent fields $\Phi(x)$
and $\Phi^*(x)$ at each space point $x$,
the inverse, $\Delta^{-1}(x-y)$,
is understood in a functional sense,
$\int d^{\,2}z\Delta^{-1}(x-z)\Delta(z-y)
=\delta^{(2)}(x-y)$,
$C_0$ is the normalization constant
unessential for the following.

We assume that the three-dimensional
nuclear densities are reduced to the product of one-nucleon
densities,
$$
\rho_N(r_1,\ldots,r_N)\,=\,\prod_{i=1}^N \rho_N(r_i),~~~~
\int d^3r \rho_N(r)=1,
$$
so that
$$
\langle\,N\,|\prod\limits_{i}F(r_i)|\,N\,\rangle\,=\,
\left[\int d^3r\,F(r)\rho_N(r)\right]^N
$$
for any function $F(r)$.

Combining the formulas (\ref{sab}) and (\ref{GI})
one gets
\begin{eqnarray}
\label{sABint}
S_{AB}(b)\,&=&\,
C_0
\int \frac{D\Phi D\Phi^*}{2\pi i}\exp\left\{
-\int d^{\,2}x d^{\,2}y\,\Phi(x)\Delta^{-1}(x-y)\Phi^*(y)
\right\} \nonumber \\
&&\times
\left[\int d^{\,2}x\,\rho_A^\bot(x-b)e^{\Phi(x)}\right]^A
\left[\int d^{\,2}y\,\rho_B^\bot(y)e^{\Phi^*(y)}\right]^B,
\end{eqnarray}
where
$$
\rho_{A,B}^\bot(x)\,=\,\int dz\,\rho_{A,B}(z,x),~~~~
\int d^{\,2}x \rho_{A,B}^\bot(x)=1
$$
are the transverse densities of the colliding nuclei $A$ and $B$.

An efficient way to deal with the integral (\ref{sABint})
is through the generating function,
\begin{eqnarray}
\label{Z}
Z(u,v)\,&=&\,
\int \frac{D\Phi D\Phi^*}{2\pi i}\exp\left\{
-\int\! d^{\,2}x d^{\,2}y\,\Phi(x)\Delta^{-1}(x-y)\Phi^*(y)
\right. \\
&&\,\left.+u\int d^{\,2}x\,\rho_A^\bot(x-b)e^{\Phi(x)} +v\int\!
d^{\,2}x\,\rho_B^\bot(x)e^{\Phi^*(x)} \right\},\nonumber \\
\label{ddZ}
S_{AB}(b)\,&=&\,\frac 1{Z(0,0)}
\frac{\partial^A}{\partial u^A}
\frac{\partial^B}{\partial v^B}\,
Z(u,v)\biggl|_{u=v=0}.
\end{eqnarray}
The short distance nature of the nuclear forces
turns the generating function into the product
of the independent integrals at the points $x_i$,
\begin{eqnarray}
\label{ZPhi}
Z(u,v)\,&=&\,
\prod\limits_{x_i}\int \frac{d\Phi(x_i) d\Phi^*(x_i)}{2\pi i}
\exp\bigl\{
-\,\frac 1{y}\,
\Phi(x_i)\Phi^*(x_i)  \nonumber \\
&&\, +\,u\,a^2\rho_A^\bot(x_i-b)e^{\Phi(x_i)}
+v\,a^2\,\rho_B^\bot(x_i)e^{\Phi^*(x_i)}
\bigr\},~~
z_y\,=\,e^{\,y}\,=\,1-\frac 12 \frac{\sigma_{NN}^{tot}}{a^2}
\nonumber
\end{eqnarray}
with the parameters $\sigma_{NN}^{tot}$ and $a$
being defined in (\ref{fNN}).
Each integral is then evaluated with the help of
the identity
\begin{equation}
\label{id}
\int \frac{d\Phi d\Phi^*}{2\pi i}\,e^{-\frac 1y\Phi \Phi^*}\,
\exp\bigl\{u e^{\Phi}\,+\,v e^{\Phi^*}\bigr\}
\,=\,
y\sum_{M,N}\frac{e^{y\,M\cdot N}}{M!N!}u^M\,v^N,
\end{equation}
resulting into (see \cite{Shabelski:2021iqk} for details)
\begin{eqnarray}
\label{Zuv}
Z(u,v)\,&=&\,e^{W(u,v)}, \\
W(u,v)\,&=&\, \frac 1{a^2}\int d^{\,2}x\,
\ln\bigl(\!\!
\sum\limits_{M\le A,N\le B}
\frac{z_y^{M\, N}}{M!N!}
\bigl[a^2 u\rho_A^\bot(x-b)\bigr]^M
\bigl[a^2 v\rho_B^\bot(x)\bigr]^N
\bigr).
\end{eqnarray}

Now we are going to apply the same method to evaluate
$J_{AB}(b)$ function (\ref{JAB}). It is the product
of two structures like (\ref{sab}) that is why
the analog of the formula (\ref{sABint}) comprises
two integrals,
\begin{eqnarray}
\label{JABint}
&&J_{AB}(b)\,=\,
C_0
\int \frac{D\Phi D\Phi^*}{2\pi i}
\int \frac{D\Psi D\Psi^*}{2\pi i} \nonumber \\
&&\times \exp\left\{
-\int d^{\,2}x d^{\,2}y\,\Phi(x)\Delta^{-1}(x-y)\Phi^*(y)
-\int d^{\,2}x^\prime d^{\,2}y^\prime\,
\Phi(x^\prime)\Delta^{-1}(x^\prime-y^\prime)\Phi^*(y^\prime)
\right\} \nonumber \\
&&\times
\langle\,A\,|\prod\limits_{i}e^{\Phi(x_i)}|\,A\,\rangle
\langle\,A\,|\prod\limits_{i}e^{\Psi(x_i^\prime)}|\,A\,\rangle
\sum\limits_{B^\prime}
\langle\,B\,|\prod\limits_{i}e^{\Phi(y_i)}|\,B^\prime\,\rangle
\langle\,B^\prime\,|\prod\limits_{i}e^{\Psi(y_i^\prime)}|\,B\,\rangle.
\nonumber
\end{eqnarray}
Recalling the completeness (\ref{complete}) one gets
\begin{eqnarray}
&&J_{AB}(b)\,=\,
C_0
\int \frac{D\Phi D\Phi^*}{2\pi i}
\int \frac{D\Psi D\Psi^*}{2\pi i} \nonumber \\
&&\times \exp\left\{
-\int d^{\,2}x d^{\,2}y\,\Phi(x)\Delta^{-1}(x-y)\Phi^*(y)
-\int d^{\,2}x^\prime d^{\,2}y^\prime\,
\Phi(x^\prime)\Delta^{-1}(x^\prime-y^\prime)\Phi^*(y^\prime)
\right\} \nonumber \\
&&\times
\left[\int d^{\,2}x\,\rho_A^\bot(x-b)e^{\Phi(x)}\right]^A
\left[\int d^{\,2}x^\prime\,\rho_A^\bot(x^\prime-b)e^{\Psi(x^\prime)}\right]^A
\left[\int d^{\,2}y\,\rho_B^\bot(y)e^{\Phi^*(y)+\Phi^*(y)}\right]^B.
\nonumber
\end{eqnarray}
Passing to the generating function we have
\begin{eqnarray}
\label{ZJ}
&&Z_J(u_A,v_A,v_B)\,=\,
\int \frac{D\Phi D\Phi^*}{2\pi i}
\int \frac{D\Psi D\Psi^*}{2\pi i} \\
&&\times \exp\left\{
-\int d^{\,2}x d^{\,2}y\,\Phi(x)\Delta^{-1}(x-y)\Phi^*(y)
-\int d^{\,2}x^\prime d^{\,2}y^\prime\,
\Phi(x^\prime)\Delta^{-1}(x^\prime-y^\prime)\Phi^*(y^\prime)
\right.
\nonumber \\
&&\,\left.+u_A\int d^{\,2}x\,\rho_A^\bot(x-b)e^{\Phi(x)} +v_A\int\!
d^{\,2}x\,\rho_A^\bot(x)e^{\Psi(x)}
+v_B\int\!
d^{\,2}x\,\rho_B^\bot(x)e^{\Phi^*(x)+\Psi^*(x)} \right\}.
\nonumber
\end{eqnarray}
For the short range interaction the integrals
over $\Phi$ and $\Psi$ variables turn into the products
of the independent integrals at the points $x_i$,
\begin{eqnarray}
\label{ZJxi}
&&Z_J(u_A,v_A,v_B)\,=\,
\prod\limits_{x_i}\int \frac{d\Phi(x_i) d\Phi^*(x_i)}{2\pi i}
\int \frac{d\Psi(x_i) d\Psi^*(x_i)}{2\pi i} \nonumber \\
&&\times \exp\bigl\{
-\,\frac 1{y}\,
\Phi(x_i)\Phi^*(x_i)
-\,\frac 1{y}\,
\Psi(x_i)\Psi^*(x_i)
\nonumber \\
&&\, +\,u_A\,a^2\rho_A^\bot(x_i-b)e^{\Phi(x_i)}
\, +\,v_A\,a^2\rho_A^\bot(x_i-b)e^{\Psi(x_i)}
+v_B\,a^2\,\rho_B^\bot(x_i)e^{\Phi^*(x_i)+\Psi^*(x_i)}
\bigr\}.
\nonumber
\end{eqnarray}
Using again the identity (\ref{id}) to evaluate the integrals
over $\Phi(x_i)$, $\Psi(x_i)$
we arrive at the generating function,
\begin{eqnarray}
\label{ZJuv}
Z_J(u_A,v_A,v_B)\,&=&\,e^{W_J(u_A,v_A,v_B)}, \\
\label{WJ}
W_J(u_A,v_A,v_B)\,&=&\, \frac 1{a^2}\int d^{\,2}x\,
\ln\bigl(\!\!
\sum\limits_{L\le 2A,K\le B}
\frac{z_y^{L\, K}}{M!N!}
\bigl[a^2 (u_A+v_A)\rho_A^\bot(x-b)\bigr]^L
\bigl[a^2 v_B\rho_B^\bot(x)\bigr]^K
\bigr). \nonumber
\end{eqnarray}
Comparing this expression with (\ref{Zuv}) we conclude
that
\begin{equation}
\label{ZJZ}
Z_J(u_A,v_A,v_B)\,=\,Z(u_A+v_A,v_B)
\end{equation}
and, respectively,
$$
J_{AB}(b)\,=\,
\frac{\partial^A}{\partial u_A^A}
\frac{\partial^A}{\partial v_A^A}
\frac{\partial^B}{\partial v_B^B}\,
Z(u_A+v_A,v_B)\biggl|_{u_A=v_A=v_B=0},
$$
or, finally
\begin{equation}
\label{d2Z}
J_{AB}(b)\,=\,
\frac{\partial^{\,2A}}{\partial u^{2A}}
\frac{\partial^B}{\partial v}\,
Z(u,v)\biggl|_{u=v=0}.
\end{equation}


\section{Results of the calculations}

The function $W(u,v)$ (\ref{Zuv}) goes as the series
built of the densities overlaps,
\begin{equation}
\label{tmn}
t_{m,n}(b)=
\frac 1{a^2}\int d^{\,2}x\,
\bigl[a^2\rho_A^\bot(x-b)\bigr]^m\,
\bigl[a^2\rho_B^\bot(x)\bigr]^n
\end{equation}
with $m\le 2A$ and $n\le B$.
For the following calculations
the nucleon density
has been taken in a simple
Gaussian parameterizations well suited for light nuclei,
\begin{equation}
\label{Gauss}
\rho(r)\,=\,\rho_0\,e^{-\frac{r^2}{a_c^2}},
\end{equation}
the value $a_c$ being expressed through the mean square
nuclear radius, $a_c=\sqrt(3/2)R_{rms}$.
The total nucleon-nucleon cross section
and the slope value (averaged over $pp$ and $pn$
interaction) are taken in the amplitude (\ref{fNN})
as~\cite{Alkhazovi:2011ty,Horiuchi:2006ga}
\begin{equation}
\label{param}
\sigma_{NN}^{tot}=43~\mathrm{mb},~~~
\beta = 0.2~\mathrm{fm}^2
\end{equation}
for the energy around 1000~MeV per projectile
nucleon.

The mean square radius $R_{ms}$ has been adjusted to match
the experimental interaction cross section
$\sigma_{^{12}\mathrm{C}-^{12}\mathrm{C}}^r=853\pm 6$~mb
at the energy about 1~GeV
per nucleon taken from the review ~\cite{Ozawa:2001hb}.
With these parameters and the overlap functions (\ref{tmn})
evaluated for the distribution (\ref{Gauss})
one gets the generating function and the amplitudes (\ref{ddZ})
and (\ref{d2Z}).

The calculations have been carried out for $\alpha$--$^{12}$C,
$^{12}$C--$^{12}$C and $^{11}$Li--$^{12}$C scattering.
The last case provides a remarkable example of a halo
nucleus $^{11}$Li, which can be treated as a core $^{9}$Li
surrounded with a halo made up of two neutrons.
The density of this composite system is assumed to be
the sum
\begin{equation}
\label{halo}
\rho(r)\,=\,N_c\rho_c(r)\,+\,N_v\rho_v(r)
\end{equation}
of the core including $N_c=9$ nucleons and
the halo with $N_v=2$ valence nucleons.
Both the densities are taken in Gaussian
form~(\ref{Gauss}), with the parameters $a_c$
being expressed through the mean square radii
of the core and the halo.
The core radius is found by calculating
the interaction cross sections of the scattering
of the nucleus $^{9}$Li, representing the core,
on the target $^{12}$C.
Comparing the output with the experimental cross sections
collected in Ref.~\cite{Ozawa:2001hb},
we tune the $R_c$ value.
Plugging it then into the density (\ref{halo})
(normalized to unity)
and comparing the calculated interaction cross sections
of the composite $^{11}$Li nucleus scattering on the $^{12}$C
with the same data set we extract the halo radius $R_v$.

Even though the table below shows the small difference between
the reaction and the interaction cross sections,
it has to be taken into account for the correct analysis of
the experimental data.

Table. Mean square radii
extracted by comparing the interaction cross section
evaluated for the given nucleus scattering on the $^{12}$C target
with the experimental cross section. The last column presents
the reaction cross sections resulting from the obtained radii.
The $^{11}$Li core radius is chosen
as that for $^{9}$Li nucleus.
The halo radius for the $^{11}$Li is found to match the interaction
cross section of its scattering on the $^{12}$C.
The experimental data are
taken from \cite{Ozawa:2001hb}
for 790~MeV.
\footnote{
The effects due to
energy dependence of the measured nucleus-nucleus interaction cross section
as well as the nucleon-nucleon scattering parameters
in the interval 790-1000~MeV
give the second order effect to the difference
$\sigma^{R}-\sigma^{I}$}

\begin{table}[H]
\centering
\begin{tabular}{|c|c|c|c|}
\hline
 & Experimental & Mean square & Reaction\\
 & cross section, mb & radius, fm & cross section, mb\\
\hline
$^{4}$He & 503 $\pm$ 5 & 1.64 & 523 \\
\hline
$^{12}$C & 853 $\pm$ 6 & 2.46 & 864 \\
\hline
$^{9}$Li & 796 $\pm$ 6 & 2.55 & 804 \\
\hline
$^{11}$Li & 1047 $\pm$ 40 & 3.28 & 1057 \\
\hline
\end{tabular}
\end{table}

The nuclear radii found from the data are to be renormalized
for the difference between the interaction cross section,
which is actually measured, and the reaction one (\ref{interact}).
The new value of $^{12}$C radius is 0.02~fm larger than that
obtained in the same Gaussian parametrization
(\ref{Gauss}) but when the reaction cross section
is matched.
This new target radius is used to get the renormalized
radii for the beam nuclei presented in the Table.
The mean square radius, $R_m$, of the $^{11}$Li,
treated as the composite core plus halo system
in the parametrization (\ref{halo}),
is expressed
through the mean square radii of the core, $R_c$ and the halo,
$R_v$, as $R_m^2=(N_c R_c^2+N_v R_v^2)/(N_c+N_v)$.
The halo radius is found to be $R_v=5.48$~fm.
The radii of the $^{9}$Li and the $^{11}$Li nuclei
are 0.11~fm and 0.13~fm larger as compared to those
obtained through the reaction cross section.

The increase of the radius when going
from the interaction to the reaction cross section
is natural since $\sigma_{AB}^{I}< \sigma_{AB}^{R}$,
although the difference between
the two cross sections in the Table
varies from 4\% for $^4$He to 1--1.5\% for more heavy nuclear beams
in a qualitative agreement with
\cite{Novikov:2013zdw}.

\section{Conclusion}

The difference between the reaction and the interaction cross
sections have been calculated for the beam nuclei $^4$He, $^{11}$Li, $^{12}$C
scattering on the $^{12}$C target.
The results are presented in the Table above.
The difference between the two values obtained
for the mean square $^{11}$Li halo radius,
which are extracted by comparing the evaluated interaction and the reaction
cross sections with the experimental one, is 0.13fm, that is about 2\%.

It is worth pointing out that the core -- halo structure
may be more complex than that underlying the density (\ref{halo}).
There could be, in principle, a state with the core $^{9}$Li
and two neutrons halo moving around their common center of mass.
However the bound state of two neutrons does not exist, which makes
three-body configuration of the neutrons and the core more subtle.


\end{document}